\documentclass[aps,prl,twocolumn,showpacs,superscriptaddress,groupedaddress,floatfix]{revtex4}  

\usepackage{rotating}

\usepackage{natbib}
\usepackage{amsmath,amssymb}
\usepackage{graphicx}
\usepackage{dcolumn}
\usepackage{bm}

\newcommand{\beq}{\begin{equation}}
\newcommand{\eeq}{\end{equation}}

\newcommand{\veczp}{\mathbf{z}^+}

\newcommand{\veczm}{\mathbf{z}^-}
\newcommand{\veczpm}{\mathbf{z}^\pm}

\begin{document}
\title{Nonlinear Interactions in Spherically Polarized Alfv\'{e}nic Turbulence}
\begin{abstract}
  Turbulent magnetic field fluctuations observed in the solar wind often maintain a constant magnitude condition accompanied by spherically polarized velocity fluctuations; these signatures are characteristic of large-amplitude Alfv\'{e}n waves. Nonlinear energy transfer in Alfv\'{e}nic turbulence is typically considered in the small-amplitude limit where the constant magnitude condition may be neglected; in contrast, nonlinear energy transfer in the large-amplitude limit remains relatively unstudied. We develop a method to analyze finite-amplitude turbulence through studying fluctuations as constant magnitude rotations in the stationary wave (de Hoffmann-Teller) frame, which reveals that signatures of finite-amplitude effects exist deep into the MHD range. While the dominant fluctuations are consistent with spherically-polarized large-amplitude Alfv\'{e}n waves, the subdominant mode is relatively compressible. Signatures of nonlinear interaction between the finite-amplitude spherically polarized mode with the subdominant population reveal highly aligned transverse components. In theoretical models of Alfv\'{e}nic turbulence, alignment is thought to reduce nonlinearity; our observations require that alignment is sufficient to either reduce shear nonlinearity such that non-Alfv\'{e}nic interactions may be responsible for energy transfer in spherically polarized states, or that counter-propagating fluctuations maintain anomalous coherence, which is a predicted signature of reflection-driven turbuence.
\end{abstract}

\author{Trevor A. Bowen}\email{tbowen@berkeley.edu}
\affiliation{Space Sciences Laboratory, University of California, Berkeley, CA 94720-7450, USA}
\author{Samuel T. Badman}
\affiliation{Space Sciences Laboratory, University of California, Berkeley, CA 94720-7450, USA}
\author{Stuart D. Bale}
\affiliation{Space Sciences Laboratory, University of California, Berkeley, CA 94720-7450, USA}
\affiliation{Physics Department, University of California, Berkeley, CA 94720-7300, USA}
\author{Thierry {Dudok de Wit}}
\affiliation{LPC2E, CNRS, CNES, and University of Orl\'eans, Orl\'eans, France}
\author{Timothy S. Horbury}
\affiliation{The Blackett Laboratory, Imperial College London, London, SW7 2AZ, UK}
\author{Kristopher G. Klein}
\affiliation{Department of Planetary Sciences \& Lunar and Planetary Laboratory, University of Arizona, Tucson, AZ 85721}
\author{Davin Larson}
\affiliation{Space Sciences Laboratory, University of California, Berkeley, CA 94720-7450, USA}
\author{Alfred Mallet}
\affiliation{Space Sciences Laboratory, University of California, Berkeley, CA 94720-7450, USA}
\author{Lorenzo Matteini}
\affiliation{The Blackett Laboratory, Imperial College London, London, SW7 2AZ, UK}
\author{Michael D. McManus}
\affiliation{Space Sciences Laboratory, University of California, Berkeley, CA 94720-7450, USA}
\affiliation{Physics Department, University of California, Berkeley, CA 94720-7300, USA}

\author{Jonathan Squire}
\affiliation{Department of Physics, University of Otago, 730 Cumberland St., Dunedin 9016, New
Zealand}
\maketitle

\paragraph{Introduction} 
Spacecraft observations of the solar wind provide much of the context for our understanding of nonlinear interactions occurring in plasma environments \cite{BrunoCarboneLRSP}. Although the solar wind is collisionless, many of its dynamics can be understood using fluid approximations, such as magnetohydrodynamics (MHD) \cite{Kulsrud1983}. To understand nonlinear interactions in MHD, the velocity $\mathbf{v}$ and magnetic field $\mathbf{B}$ are cast into the Elsasser variables
${\bf{z}}^\pm={\bf{v}} \pm  {{\bf{b}}}$ with
${\bf{b}}={{\bf{B}}}/{\sqrt{\mu_0\rho}}$,
and mass density $\rho$ \cite{Elsasser1950}.
Assuming incompressbility, which is motivated by our observations,
\begin{align}\label{eq:inc_mom}
\partial_{t}{\bf{z}}^\pm=-{\bf{z}}^\mp\cdot \nabla {\bf{z}}^\pm -\frac{1}{\rho}\nabla (p_{th}+ \frac{\rho}{2} b^2). 
\end{align}
The Elsasser fluctuations can be identified as travelling parallel ($\veczm$) and anti-parallel ($\veczp$) to the mean magnetic field $\mathbf{b}_0$\cite{Alfven1942, Elsasser1950}. Fluctuations in the solar wind,  defined as $\delta{\bf{x}}={\bf{x}}-{\bf{x}}_0$, are dominated by fluctuations perpendicular to $\mathbf{b}_0$ with $\delta \mathbf{v_\perp} \approx \pm \delta \mathbf{b_\perp}$  \cite{Belcher1971}. In the small-amplitude limit, Alfv\'{e}n waves are approximated by transverse linear plane-waves polarized perpendicular to both wave-vector $\mathbf{k}$ and the background field $\mathbf{b}_0$. This limit is well described by Reduced MHD (RMHD), where $\mathbf{z}^{\pm}$ perturbations are perpendicular, $\mathbf{z}^{\pm}=\mathbf{z}^{\pm}_{\perp}$, polarized perpendicular
to their wave vector ${\bf k}$, and can be written using the Elsasser potentials $\zeta^{\pm}$ as $\mathbf{z}^{\pm}=\hat{\bf b}_{0}\times \nabla_{\perp}\zeta^{\pm}$ \cite{Strauss1976,Schekochihin2009}. Observations in the solar wind are often dominated by one of $\delta\veczpm_\perp$\cite{Belcher1971}--a condition that is commonly known as imbalance \cite{Lithwick2007}. In the solar wind, imbalance favors the anti-sunward propagating Alfv\'{e}n wave population \cite{Roberts1987a,Roberts1987b}. 
In turn, the subdominant Elsasser variable corresponds to a sunward propagating mode that interacts nonlinearly with the dominant mode via the ${\delta\bf{z}}^\mp\cdot \nabla {\delta\bf{z}}^\pm$ term in Equation \ref{eq:inc_mom}. This mutual shearing of counter-propagating fluctuations in MHD is thought to generate a cascade, similar to neutral-fluid turbulence, that results in inertial range energy transfer \cite{Kolmogorov1941,Kraichnan1965,Coleman1968,Dobrowolny1980,Grappin1982,GS95,Boldyrev2005,Boldyrev2006,Mason2006,Schekochihin2009,Chandran2015,Howes2015}.

Theoretical consideration of magnetized turbulence is often performed in a small-amplitude limit \cite{Kolmogorov1941,Kraichnan1965,Coleman1968,Strauss1976,Grappin1982,Dobrowolny1980,GS95,Boldyrev2006,Schekochihin2009,Boldyrev2005,Beresnyak2006,Mason2006,Podesta2009,Chandran2015,Mallet2015,Howes2015,Mallet2016}; however, the solar wind is subject to large-amplitude ($|\delta \mathbf{b}|/ |\mathbf{b}_0|\sim1$) fluctuations that maintain  $$|\mathbf{b}|=|\mathbf{b}_0+\delta \mathbf{b}|=const,$$ and thus appear spherically-polarized \cite{Goldstein1974,Lichtenstein1980,Riley1996}. This condition is characteristic of large-amplitude Alfv\'{e}n waves, which are not entirely perpendicular to $\mathbf{b}_0$, but can acquire a component parallel to $\mathbf{b}_0$ in order to maintain constant magnitude of the total magnetic field \cite{BarnesHollweg1974,Goldstein1974}.  The small-amplitude limit neglects the higher-order corrections that maintain constant total magnitude. 

Observations from Parker Solar Probe (PSP) reveal that both the constant-magnitude condition and high-Alfv\'{e}nicity are pronounced in the inner-heliosphere \cite{Bale2019,DudokdeWit2020,Horbury2020,Chen2020,McManus2020,Chaston2020,Bourouaine2020,Chen2021,Martinovic2021}, and are consistent with the large-amplitude Alfv\'{e}n mode; however, these studies mostly omit discussion of observed finite-amplitude signatures in turbulence. 
Both  magnetic  field and  velocity  fluctuations show finite-amplitude signatures of spherical polarization \cite{Wang2012,Matteini2015} consistent with the transverse Alfv\'{e}n mode \cite{BarnesHollweg1974}. While spherical polarization of the velocity fluctuations is independent of (Galilean) reference frame, constant-magnitude of the total velocity vector is only maintained in the frame at the center of spherical polarization \cite{BarnesHollweg1974,Matteini2015}, associated with the de Hoffmann-Teller frame (dHTf) \cite{deHoffmannTeller1950}. 
Analysis in the dHTf enables fluctuations to be characterized in terms of constant magnitude \emph{rotations} corresponding to large-amplitude Alfv\'{e}n waves \cite{BarnesHollweg1974,Hollweg1974}. 
Formally, the dHTf minimizes the convected electric field ${\bf{E}}=-{\bf{v}} \times {\bf{B}}$, which is ideally zero in a stationary frame of transverse electromagnetic waves. The existence of an empirically measurable dHTf indicates significant {\em{alignment}} between magnetic and velocity fluctuations, which constrains turbulent energy transfer.  

In this Letter, we discuss the nonlinear interactions observed in spherically polarized turbulence, which we measure to be highly aligned in nature. The relation between alignment and nonlinearity is clear when RMHD is written using the Elsasser potentials 
\begin{equation}\partial_{t}\nabla_{\perp}^{2}\zeta^{\pm}
\propto \left\{\zeta^{+}, \nabla_{\perp}^{2} \zeta^{-}\right\}+\left\{\zeta^{-}, \nabla_{\perp}^{2} \zeta^{+}\right\} \mp \nabla_{\perp}^{2}\left\{\zeta^{+}, \zeta^{-}\right\},\end{equation}
with $\{A,B\}=\left(\nabla_{\perp} A \times \nabla_{\perp} B\right)\cdot \hat{{\mathbf{b}}}_{0}.$ Because $\{A,B\}$ vanishes if $\nabla_{\perp}A$ and $\nabla_{\perp}B$ are parallel, 
the final term vanishes for aligned $\mathbf{z}_{\perp}^{\pm}$. 
The other nonlinear terms are significantly reduced if contours of  $\nabla_{\perp}^{2}\zeta^{\pm}$ are approximately aligned 
with $\zeta^{\mp}$, a condition that is satisfied for perturbations that look locally like sheets or tubes. 
In the theory of dynamic alignment, cascading turbulent structures are sheet-like and $\delta \mathbf{z}^{+}$ and $\delta \mathbf{z}^{-}$ shear themselves into alignment (i.e., becoming more parallel at smaller scales), such that $\delta \mathbf{z}_{\perp}^{\mp} \cdot \nabla \delta \mathbf{z}_{\perp}^{\pm}$ is reduced by a factor $\sim\!\sin\phi_{\bf k}$, where $\phi_{\bf k}$ is the alignment angle between $\delta \mathbf{z}_{\perp}^{+}$ and $\delta \mathbf{z}_{\perp}^{-}$ at scale ${\bf k}$. This depletes the nonlinearity towards smaller scales, flattening $\sim k_{\perp}^{-5/3}$, energy spectrum to $\sim k_{\perp}^{-3/2}$, which is often observed in the solar wind \cite{Podesta2009,Chen2013,Bowen2018b,Chen2020}. 


The dynamic alignment argument neglects the fact that perfectly aligned fluctuations may have significant nonlinearity. Because $\delta \mathbf{z}^{\pm} \cdot \nabla \delta \mathbf{z}^{\mp} = \delta \mathbf{z}^{\mp} \cdot \nabla \delta \mathbf{z}^{\pm} -\nabla\times ( \delta \mathbf{z}^{\pm}\times \delta \mathbf{z}^{\mp}) $ it is possible to have both $\mathbf{z}_{\perp}^{+}\propto\mathbf{z}_{\perp}^{-}$ and significant nonlinearity 
as long as $\delta \mathbf{z}^{\mp} \cdot \nabla \delta \mathbf{z}^{\pm} \approx \delta \mathbf{z}^{\pm} \cdot \nabla \delta \mathbf{z}^{\mp}$. 
In essence, it is possible that $\mathbf{z}_{\perp}^{+}\propto\mathbf{z}_{\perp}^{-}$ ($\zeta^{+}\propto\zeta^{-}$), while not satisfying 
a specific topology, in which case aligned fluctuations have significant nonlinearity.
Here, we show that measurements of highly aligned $ \mathbf{z}^{\pm}$ suggests either (i) that large-amplitude fluctuations are sheet or tube-like structures and nonlinearity is significantly reduced to an extent that non-Alfv\'{e}nic modes contribute significantly to nonlinear energy transfer, or (ii) 
$ \mathbf{z}^{-}$ and $\mathbf{z}^{+}$ are ``anomalously coherent'' (nearly proportional), but that alignment does not affect nonlinear energy transport. 

\begin{figure}[!]
    \centering
    \includegraphics[width=2.5in]{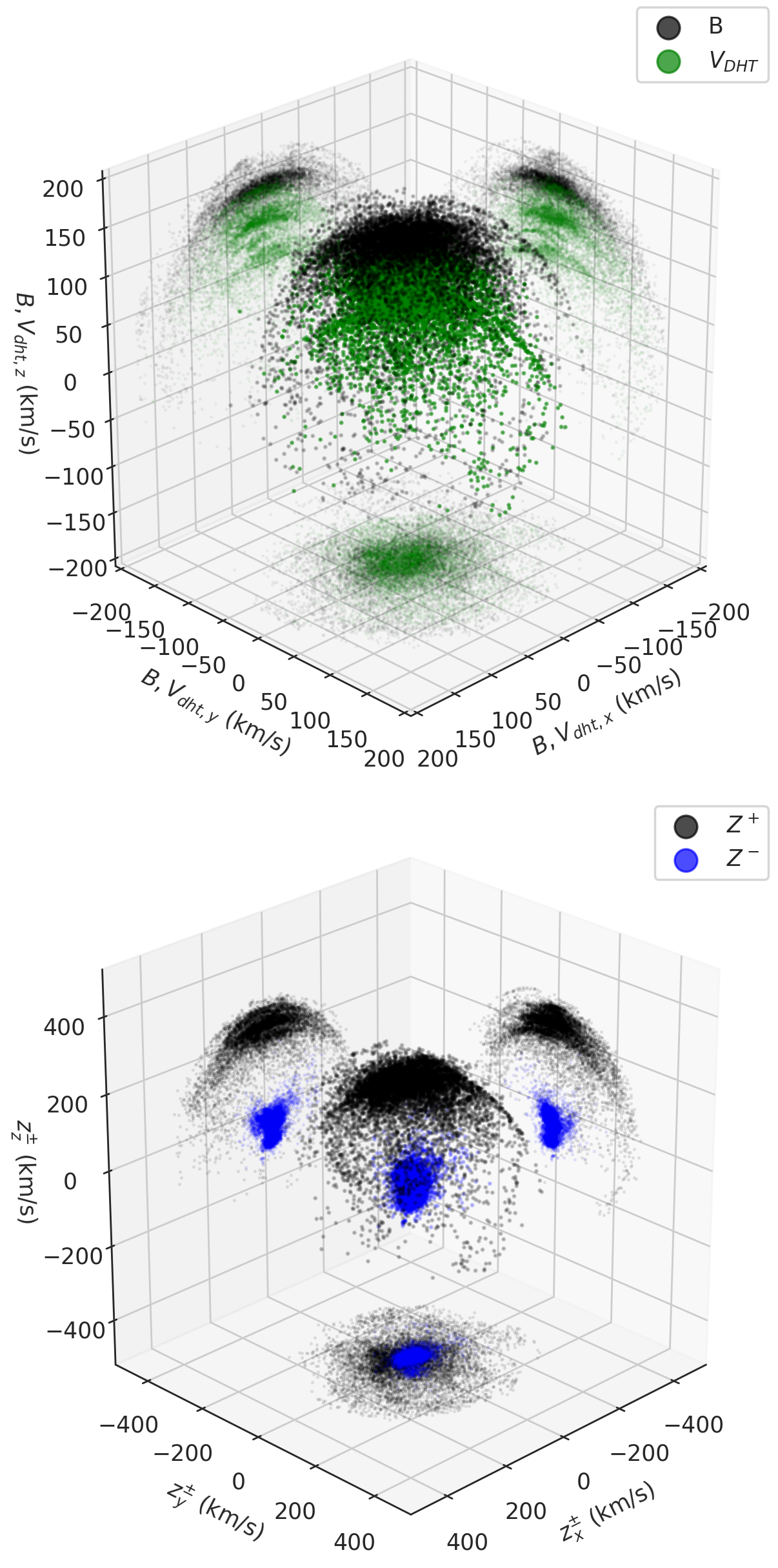}
   \caption{(a) Velocity measurements in dHTf (green), magnetic field fluctuations in Alfv\'{e}n units (black). (b) Elsasser variables $\mathbf{z}'\pm$ constructed in the dHTf (black +, blue -).}
    \label{fig:1}
\end{figure}

\paragraph{Data}
PSP provides measurements of the inner heliosphere using the electromagnetic FIELDS \citep{Bale2016} and Solar Wind Electron Alpha and Proton (SWEAP, \cite{Kasper2016}) instrument suites. We study a stream from PSP perihelion 4 from 2020-01-28/09:56-21:43. The mean magnetic field was directed sunward, such that ${\bf{z}^+}$ is the outward propagating mode. Magnetic field data is obtained from PSP/FIELDS \cite{Bale2019,Bowen2020b}. Nonlinear bi-Maxwellian fits to proton beam and core distributions measured by the PSP/SWEAP Solar Probe ANalyzer (SPANi) provide estimates of ${\bf{v}}$, $T_{\parallel,\perp}$ \cite{Kasper2016,Klein2021}. The proton core and beam are well resolved. Density measurements are from quasi-thermal noise (QTN) from FIELDS\cite{Malaspina2016,Pulupa2017}.
Uncertainty in the density is order 10\% due to the resolution of FIELDS/RFS \cite{Pulupa2017}.

\paragraph{Turbulent Signatures in the dHTf} Fig. \ref{fig:1}(a) shows measurements of ${\bf{b}}$ and ${{\bf{v}}'={\bf{v}}-{\bf{v}}_{dHT}}$, with projections on each 2D plane. ${\bf{v}}_{dHT}$ is found through minimizing
\begin{align}\label{eq:dht} 
E^2=\sum_i \left(({\bf{v}}_{dHT} -{\bf{v}}_i)\times {\bf{B}}_i\right)^2
\end{align}
with respect to ${\bf{v}}_{dHT}$.
Fig. \ref{fig:1}(b) shows the dHTf Elsasser variables defined as ${{\bf{z}'}^{\pm}}=\mathbf{v}' \pm{\bf{b}}$, with 
${\bf{v}}_{{dHT}}=[-71,23,-342]$ km/s $\langle \mathbf{v}\rangle={\bf{v}}_{sw}=[-99,26,-265]$ km/s. The Alfv\'{e}n speed is ${\bf{v}}_A=\langle{\bf{b}}\rangle=[19,10,97]$ km/s.  The mean solar wind speed is $\langle|{\bf{v}}_{sw}|\rangle=295$ \text{km/s}, with central half of values $265<|{\bf{v}}_{sw}|<310$ km/s. Fig. \ref{fig:1}(b) shows that ${{\bf{z}}'^+}$ is well approximated by a spherical surface, with constant magnitude $|{{\bf{z}}'^+}|$=221 km/s. In contrast, ${{\bf{z}}'^-}$ has significant compressibility. The magnetic field is taken as frame invariant and no substantial effects were found when introducing kinetic Alfv\'{e}n speed normalizations \cite{Barnes1979,Chen2013}.

A local, scale-dependent, dHTf, ${\bf{\tilde{v}}}_{t,t+\tau}$, is constructed using two point increments and averages \begin{align}{\Delta {\bf{x}}=\bf{x}}(t+\tau)-{\bf{x}}(t)\label{equation:inc}\\
\bar{\bf{x}}=\frac{{\bf{x}}(t)+{\bf{x}}(t+\tau)}{2}\label{equation:mean}\end{align}
in  $\mathbf{v}$ and $\mathbf{b}$ for every increment pair \cite{1998ISSI},

\begin{align}
E_{t,t+\tau}^2=\left[({\bf{\tilde{v}}} -{\Delta\bf{v}})\times ({\bar{ {\bf{b}}}+\Delta {\bf{b}}})\right]^2\\
\frac{\partial{E^2}}{\partial{\bf{\tilde{v}}}}=0.
\end{align}
\begin{figure}
    \centering
    \includegraphics[width=3in]{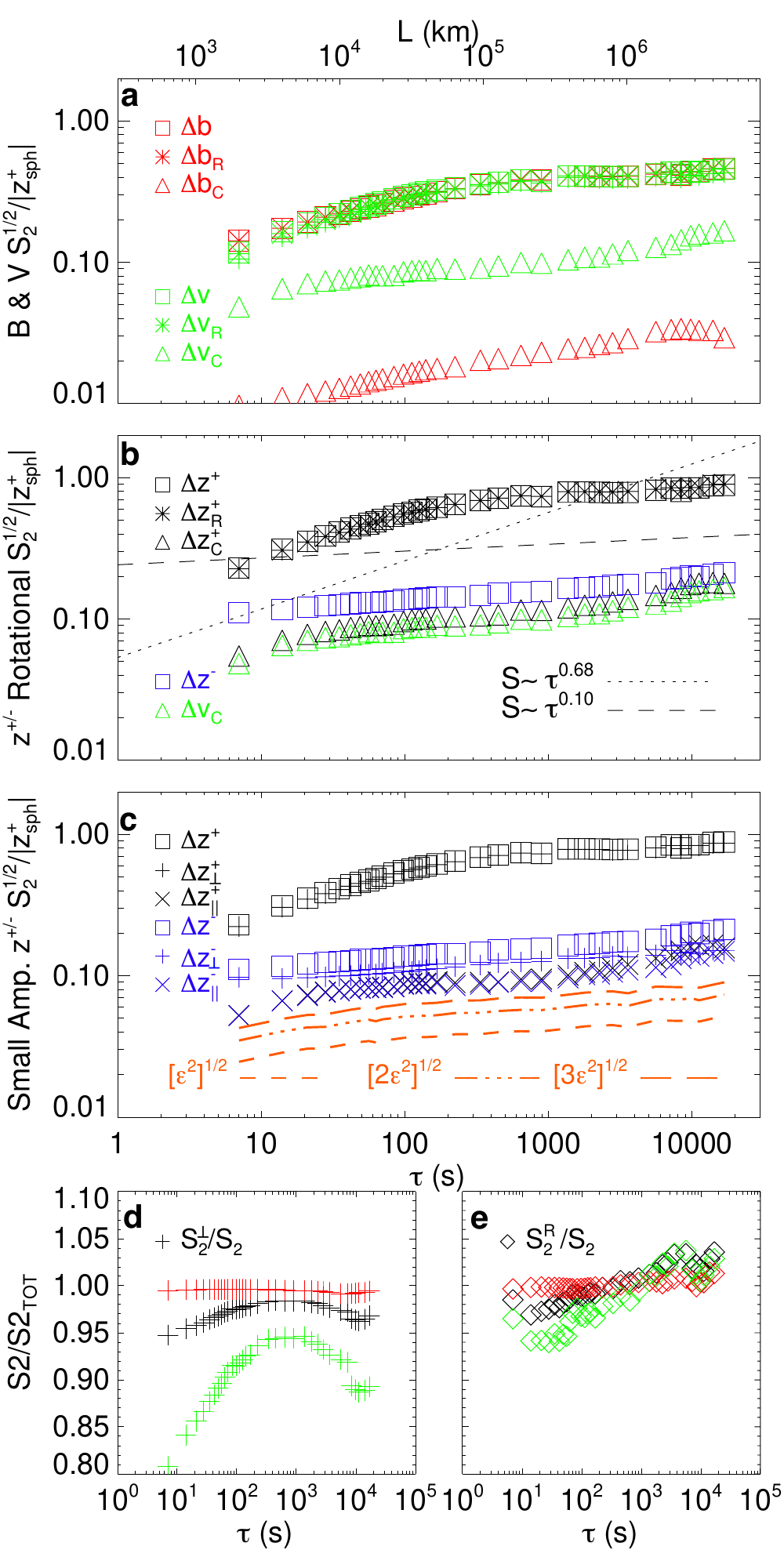}
   \caption{(a-b) Structure functions $S_2$ for finite-amplitude increments ($\Delta {\bf{x}}$, $\Delta {\bf{x}}_R$, $\Delta {\bf{x}}_C.$) for  $ {\bf{b}}$ (red), $ {\bf{v}}'$ (green) and ${\bf{z}}^+$ (black) normalized to $|\mathbf{z}'^+|$; total $\Delta{\bf{z}}^-$ is in blue. Power-law fits to structure functions of $\Delta {\bf{z}}^+$ are in black lines. (c) Small amplitude $\Delta {\bf{z}}'^+$ (black), and $ \Delta {\bf{z}}'^-$ (blue) increments.  Perpendicular $\perp$ and parallel $\parallel$ increments are shown respectively as $+$ and $\times$. As described in the text, a bound on instrumental noise determined from unaligned increments in $\mathbf{v}$ is shown in orange. Panels (d) and (e) show the ratio of  $S_2^\perp/S_2$ and $S_2^R/S_2$ for $\Delta {\bf{z}}'^+$, $\Delta {\bf{v}}'$ and $\Delta {\bf{b}}$.}
    \label{fig:2}
\end{figure}
\paragraph{Rotational Increments} 
In the small-amplitude limit, perpendicular and parallel increments discern between Alfv\'{e}nic and compressible fluctuations \cite{Mason2006,Schekochihin2009,Podesta2009,Wicks2013a,Wicks2013b}
\begin{align}
\label{eq:small_amp_par}
 \Delta {\bf{x}}_\parallel(t,\tau)=&(\Delta {\bf{x}}(t,\tau)\cdot\hat{\bar{\mathbf{b}}})\hat{\bar{\mathbf{b}}},\\
 \label{eq:small_amp_perp}
 \Delta {\bf{x}}_\perp(t,\tau)=&\Delta {\bf{x}}(t,\tau)- \Delta {\bf{x}}_\parallel(t,\tau),\\
     \hat{{\bf{x}}}=&\frac{{{\bf{x}}}}{|{\bf{x}}|}.
 \end{align}
In contrast, the finite-amplitude Alfv\'{e}n mode is associated with a constant magnitude rotation, we thus introduce a \emph{rotational increment}:

 \begin{align}
\Delta {\bf{x}}_R(t,\tau)=&{\bf{x}}(t)-{\bf{R}}_{t,t+\tau}{\bf{x}}(t),\\
    {\bf{R}}_{t,t+\tau}=& \hat{{\bf{x}}}(t+\tau)\hat{{\bf{x}}}(t)^T\end{align}
    which is the part of $\Delta \mathbf{x}$ involving only a rotation of $\mathbf{x}$. The rotational increment is only sensible to define for $\mathbf{v}$ and $\mathbf{z}^+$ fluctuations in the dHTf, where the velocity fluctuations have (approximately) constant magnitude. Outside of the dHTf, a constant magnitude rotation in velocity includes contribution from the offset-center of the spherical surface into the estimate of $\Delta\mathbf{v}_R$. Each rotational increment for $\mathbf{v}'$ and $\mathbf{z}'^+$ is defined in its local, scale dependent, ${\bf{\tilde{v}}}_{t,t+\tau}$ frame. 
    The non-rotational portion of the increment $\Delta {\bf{x}}_C$ is
    \begin{align}
    \Delta {\bf{x}}_C=\Delta {\bf{x}}-\Delta {\bf{x}}_R.
    \end{align}
    
    Fig.~\ref{fig:2}(a) shows the square-root of the trace of the second order structure functions $S_2=\langle\Delta {\bf{x}}^2\rangle$ for finite-amplitude increments in  ${\bf{b}}$ and ${\bf{v}}'$. The amplitudes are normalized to the magnitude $|\mathbf{z}'^+|=221$ km/s. The non-rotational portion of the velocity fluctuations $\Delta {\bf{v}}_C$ is significantly larger than $\Delta {\bf{b}}_C$. Fig.~\ref{fig:2}(b) shows total, rotational, and non-rotational increments in ${\bf{z}}'^\pm$ normalized to $|\mathbf{z}'^+|$. Alfv\'{e}nic fluctuations in ${\bf{z}}'^-$ are not expected to be pure rotations in the empirically measured dHT, and thus $\Delta {\bf{z}}_C^-$ and $\Delta {\bf{z}}_R^-$ are omitted. The total $\Delta{\bf{z}}^+$ increments are almost precisely equal to the rotational $\Delta{\bf{z}}_R^+$ increments, indicating that the fluctuations are dominated by the finite-amplitude Alfv\'{e}n wave. $\Delta {\bf{z}}_C^+$ is dominated by non-rotational  $\Delta {\bf{v}}_C$ fluctuations.
    
    Power-law scalings are fit to the $\Delta{\bf{z}}'^+$ increments with $S^{z+}_2\sim\tau^\alpha$. An outer range index is $\alpha=0.1$, while a steeper MHD range scaling, commonly associated with an inertial range, is found to be $\alpha=0.7$. 
    A spectral break occurs at approximately 100 s or, assuming Taylor's hypothesis, length-scales of $L\sim30000$ km. 
    
 Fig.~\ref{fig:2}(c) shows the application of small amplitude increments (Equations \ref{eq:small_amp_par} \& \ref{eq:small_amp_perp}) to the Elsasser variables. Measurable differences between the small amplitude and rotational increments in Fig.~\ref{fig:2}(d,e), which are admittedly small, are a signature of the finite-amplitude nature of the fluctuations. Fig.~\ref{fig:2}(e) shows the persistence of the constant total magnitude condition well below outer-scales, suggesting that the constant magnitude condition is dynamically relevant to nonlinear energy transfer across MHD scales \cite{Matteini2018,Matteini2019}.
 
 
 \paragraph{Nonlinearities}

\begin{figure}
    \centering
    \includegraphics[width=3in]{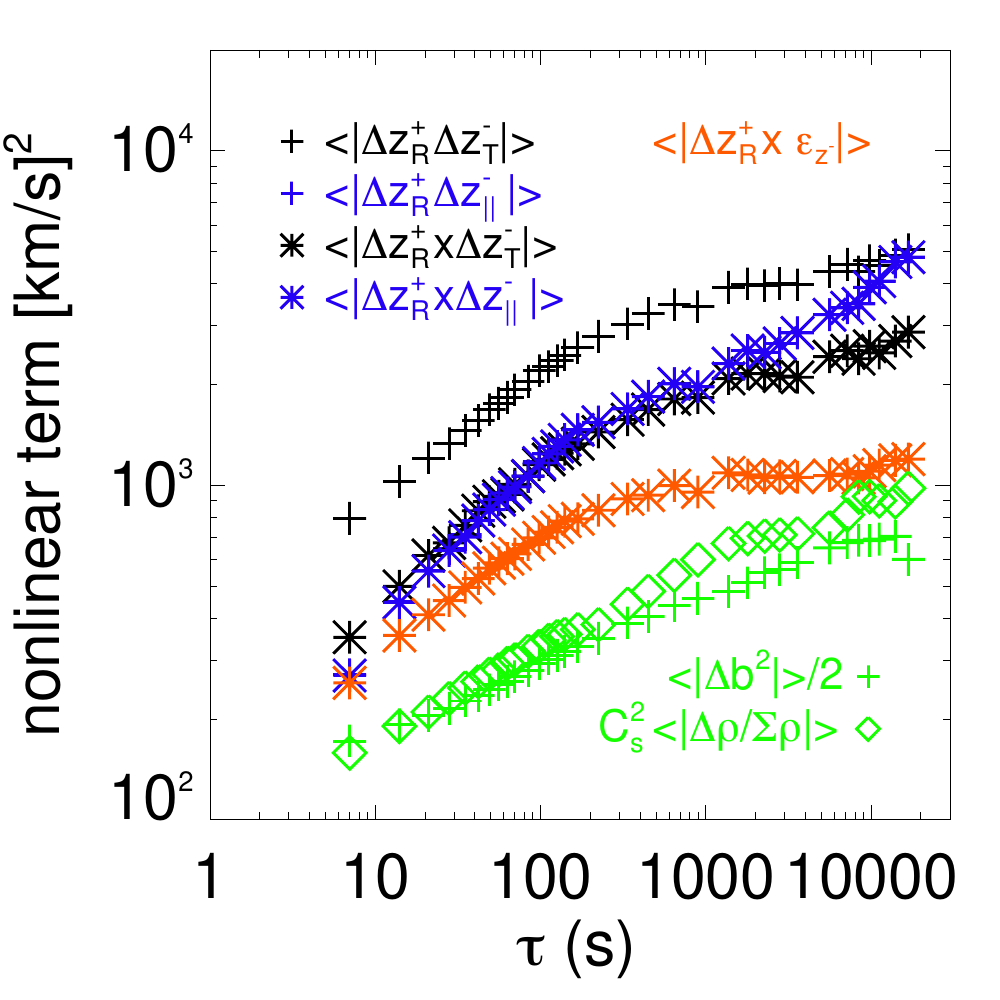}
    \caption{Cross-term structure functions $\langle|\Delta\mathbf{z}^+_R||\Delta\mathbf{z}^-_\perp|\rangle$ and $\langle|\Delta\mathbf{z}^+_R||\Delta\mathbf{z}^-_\parallel|\rangle$(black, blue +). Cross-term structure functions under consideration of alignment $\langle|\Delta\mathbf{z}^+_R\times\Delta\mathbf{z}^-_\perp|\rangle$ and $\langle|\Delta\mathbf{z}^+_R\times\Delta\mathbf{z}^-_\parallel|\rangle$ (black, blue,*).Magnetic $1/2{\Delta b^2}$ and thermal pressure  $\Delta p_{th}$ terms are shown in green + and $\diamond$. An error estimate $\epsilon_{\Delta z^-} \times \Delta{\mathbf{z}}_{R}$ described in the text is shown in orange.}
    \label{fig:3}
\end{figure}

  Nonlinear turbulent interactions of finite-amplitude Alfv\'{e}n waves are poorly understood, yet key to understanding how these fluctuations evolve and dissipate.
 We study the strength of nonlinear interactions using cross-terms  $\langle|\Delta\mathbf{x}||\Delta\mathbf{y}|\rangle$\cite{Mason2006,Beresnyak2006}; this assumes that nonlinear interactions are local in scale. The strength of the nonlinearity associated with transverse shear Alfv\'{e}n wave interactions is given as $\langle|\Delta\mathbf{z}_R^+||\Delta\mathbf{z}^-_\perp|\rangle$. 
Similarly, we estimate shear interactions not associated with Alfv\'{e}n modes using $\langle|\Delta\mathbf{z}^+_R||\Delta\mathbf{z}^-_\parallel|\rangle$.  Fig.~\ref{fig:3} shows the mean magnitude quantities of nonlinear terms. 
The transverse nonlinearity $\langle|\Delta\mathbf{z}^+_R||\Delta\mathbf{z}^-_\perp|\rangle$ dominates across MHD scales. Fig.~\ref{fig:3} additionally shows nonlinearity in the magnetic pressure $1/2\langle{\Delta b^2}\rangle$  \begin{align}\Delta b^2={{b}^2}(t+\tau)-{{b}^2}(t),\end{align} and thermal pressure gradients assuming $p_{th}=c_s^2\rho$\begin{align}
 c_s^2/b_0^2=\beta&=\frac{2\mu_0n_p}{B^2}(2/3 T_{\perp_p}+1/3 T_{\parallel p}),\\
 \Delta p_{th}=&\beta \bar{b_0^2}\frac{\Delta \rho}{\bar{\rho}},
 \end{align}
where $T_{\perp,\parallel}$ is defined using proton core and beam populations \cite{Klein2021}. The median $\beta=0.66$ with the central half of values between $0.48<\beta<0.75$. Though the isothermal equation of state is likely not well satisfied, and full compressible treatment requires additional terms in Equation \ref{eq:inc_mom}, Fig.~\ref{fig:3} shows the dominance of shear nonlinearities, justifying an incompressible approximation.
We study the effect of alignment on nonlinearity using the terms $\langle|\Delta\mathbf{z}_R^+\times\Delta\mathbf{z}^-_\perp|\rangle$ and $\langle|\Delta\mathbf{z}^+_R\times\Delta\mathbf{z}^-_\parallel|\rangle$ \cite{Beresnyak2006,Mason2006,Servidio2008,Podesta2009}. Fig.~\ref{fig:3} shows that the alignment strongly affects the shear Alfv\'{e}n-wave nonlinearity. This effect is observed across scales, such that if the shear nonlinearity is depleted according to $ {\delta\bf{z}}^\mp\cdot \nabla {\delta\bf{z}}^\pm\sim \nabla \times ({\delta\bf{z}}^\pm \times {\delta\bf{z}}^\mp)$, then the non-transverse terms become as strong as the Alfv\'{e}nic nonlinearity. Reduction in the Alfv\'{e}nic nonlinearity indicates that the rotational $\Delta\mathbf{z}_R^+$ is highly aligned with $\Delta\mathbf{z}_\perp^-$. In contrast, including alignment has little effect on the nonlinearity associated with $\mathbf{z}^-_\parallel$.



The sensitivity of our results to instrumental noise is tested by computing unaligned fluctuations in $\Delta \mathbf{v}_\perp$ as $\epsilon_v=\Delta \mathbf{v}_\perp \times \hat{\Delta \mathbf{b}_\perp}$, as a bound on measurement uncertainty (i.e. assuming any unaligned $\Delta \mathbf{v}_\perp$ is entirely noise). The rms value of $\epsilon_v$ at smallest scales ($\tau_{min}$=7s) is 5.4 km/s, which is an upper-bound on the error and corresponds to $\sqrt{\epsilon_v^2}\sim 0.02|\mathbf{z}'^+|$; $\Delta {\mathbf{z}}_\parallel \sim 0.08|\mathbf{z}'^+|$, at smallest scales.
The trace noise is $ \sqrt{3\epsilon_v^2} \sim 9$ km/s or $\sqrt{3\epsilon_v^2}\sim 0.05|\mathbf{z}'^+|$. Smallest $\Delta\mathbf{z}_\perp^-$ fluctuations are at $\sim|0.1\mathbf{z}'^+|$, Fig.~\ref{fig:2}(c). Sensitivity in the nonlinear cross-term structure functions is bounded assuming  $\Delta \mathbf{z^-}$ at the smallest scale is purely noise $\epsilon_{\Delta z^-}=\Delta \mathbf{z}_{\tau_{min}}^-$; the scale dependent noise is computed as $\epsilon_{\Delta z^-} \times \Delta{\mathbf{z}}_{R}$.

\paragraph{Conclusions}

 Analysis of MHD fluctuations in the dHTf enables the study of Alfv\'{e}nic turbulence in a large-amplitude limit, in which the dominant mode corresponds to constant magnitude rotations \cite{BarnesHollweg1974,Goldstein1974}.  The Alfv\'{e}nic constant $|B|$ state is a nonlinear equilibrium solution satisfied in physical space that affects all scales contributing significantly to $\delta B$. Our development of rotational increments shows that the constant magnitude condition, and large amplitude fluctuations $\delta B$/$B > 0.3$, can be maintained deep into scales traditionally associated with an ``inertial range'' (Fig.~\ref{fig:2}d-e) \cite{Kraichnan1965,GS95}. Our observations suggest that fluctuations across all MHD scales may maintain correlations to keep $|B|$ constant \cite{Matteini2018}.

While incompressible shearing between counter-propagating Alfv\'{e}n waves is often invoked in describing magnetized Alfv\'{e}nic turbulence, theories are typically developed in limits that omit correlations that maintain the constant magnitude condition\cite{Kraichnan1965,Coleman1968,Strauss1976,Grappin1982,Dobrowolny1980,GS95,Boldyrev2006,Beresnyak2006,Schekochihin2009,Boldyrev2005,Mason2006,Podesta2009,Chandran2015,Mallet2015,Howes2015,Mallet2016}. 
The presence of correlations between that maintain constant magnitude throughout MHD-range fluctuations, suggest that finite-amplitude dynamics \cite{Derby1978,Vinas1991,DelZanna2001,Tenerani2013,Squire2017,Bowen2018a,Chandran2018,Gonzalez2020,Tenerani2020} are likely important nonlinear processes. Finite-amplitude effects may contribute to the growth and evolution of MHD-scale compressible fluctuations, as well as kinetic processes \cite{Matteini2010, Gonzalez2021}.

Our observations show that large-amplitude fluctuations in the solar wind are highly aligned and that measured nonlinearities are not in agreement with common understandings of alignment on Alfv\'enic turbulence \cite{Boldyrev2006}. Across all scales, the Alfv\'enic (rotational) increments  of  $ {\bf z}^{+}$ are aligned with Alfv\'enic (perpendicular) increments  
of ${\bf z}^{-}$, such that $\langle|\Delta \mathbf{z}_{R}^{+} \times \Delta \mathbf{z}_{\perp}^{-}|\rangle\ll \langle|\Delta \mathbf{z}_{R}^{+} || \Delta \mathbf{z}_{\perp}^{-}|\rangle$. Strong alignment only affects the Alfv\'enic part of ${\bf z}^{-}$: the parallel ${\bf z}^{-}$ increment is not strongly 
aligned ($\langle|\Delta \mathbf{z}_{R}^{+} \times \Delta \mathbf{z}_{\|}^{-}|\rangle\approx \langle|\Delta \mathbf{z}_{R}^{+} || \Delta \mathbf{z}_{\|}^{-}|\rangle$),
such that $\langle|\Delta \mathbf{z}_{R}^{+} \times \Delta \mathbf{z}_{\perp}^{-}|\rangle \approx \langle|\Delta \mathbf{z}_{R}^{+} \times \Delta \mathbf{z}_{\|}^{-}|\rangle$
even though $\langle|\Delta \mathbf{z}_{\|}^{-}|^{2}\rangle < \langle|\Delta \mathbf{z}_{\perp}^{-}|^{2}\rangle$. The nonlinearity associated with the magnetic pressure is weak relative to the shear interactions, suggesting that the waves relax to the constant magnitude state efficiently \cite{CohenKulsrud1974}. Recent-work shows that small perturbations in magnetic pressure may be an effect of expanding large-amplitude Alfv\'{e}n waves with oblique wave-vectors \cite{Mallet2021}.

Interpreting these results hinges on the relevance of alignment to strength of nonlinear interactions. If the Alfv\'{e}nic nonlinearity is depleted by $\sim\!\sin\phi_{\bf k}$, \cite{Boldyrev2006}, then our results imply that the non-Alfv\'enic nonlinearity (involving ${\bf z}^{-}_{\|}$) is approximately equal to the Alfv\'enic shear-nonlinearity, challenging the basis for RMHD turbulence phenomenologies. 
If alignment is not directly related to the nonlinearity, which occurs  if  
$\delta \mathbf{z}^{\mp} \cdot \nabla \delta \mathbf{z}^{\pm} \approx  \delta \mathbf{z}^{\pm} \cdot \nabla \delta \mathbf{z}^{\mp} $,
then our results imply that ${\bf z}^{+}$ and ${\bf z}^{-}$ are anomalously  
coherent, despite propagating in opposite directions. Either way, our results do not support a dynamically aligned cascade phenomenology often invoked to explain observed spectra \cite{Boldyrev2005,Boldyrev2006,Mason2006}. Our results further suggest that alignment has little bearing on the development of inertial range turbulence from outer scales \cite{Wicks2013a,Wicks2013b}.
The possibility that nonlinearity is unrelated to alignment, and that the Elsasser variables are anomalously coherent, resembles theories of reflection-driven turbulence \cite{Perez2013,Velli1989,Chandran2019}, where  ${\bf z}^{-}$ is strongly aligned with ${\bf z}^{+}$ because it is driven directly by wave reflection from large-scale gradients. In any case, our results question the relevance of RMHD turbulence theories to the inner-heliosphere and provide constraints on dynamics of finite-amplitude turbulence.

\begin{acknowledgements}The authors thank Prof. Benjamin Chandran, Dr. Christopher H.K Chen,  Dr. Romain Meyrand and Dr. Julia Stawarz for useful discussions surrounding this manuscript. TAB is suppored through NASA Grant  NNN06AA01C KGK is supported by NASA Grants NNN06AA01C and 80NSSC19K0829. Support for JS was provided by Rutherford Discovery Fellowship RDF-U001804 and Marsden Fund grant UOO1727, which are managed through the Royal Society Te Ap\=arangi. \end{acknowledgements}

\providecommand{\noopsort}[1]{}\providecommand{\singleletter}[1]{#1}%


\begin{thebibliography}{78}
\providecommand{\natexlab}[1]{#1}
\providecommand{\url}[1]{\texttt{#1}}
\expandafter\ifx\csname urlstyle\endcsname\relax
  \providecommand{\doi}[1]{doi: #1}\else
  \providecommand{\doi}{doi: \begingroup \urlstyle{rm}\Url}\fi

\bibitem[{Bruno} and {Carbone}(2013)]{BrunoCarboneLRSP}
Roberto {Bruno} and Vincenzo {Carbone}.
\newblock {The Solar Wind as a Turbulence Laboratory}.
\newblock \emph{Living Reviews in Solar Physics}, 10\penalty0 (1):\penalty0 2,
  May 2013.
\newblock \doi{10.12942/lrsp-2013-2}.

\bibitem[Kulsrud(1983)]{Kulsrud1983}
R~M Kulsrud.
\newblock {MHD description of plasma}.
\newblock In R~N Sagdeev and M~N Rosenbluth, editors, \emph{{Handbook of Plasma
  Physics}}. Princeton University, 1983.

\bibitem[Elsasser(1950)]{Elsasser1950}
Walter~M. Elsasser.
\newblock The hydromagnetic equations.
\newblock \emph{Phys. Rev.}, 79:\penalty0 183--183, Jul 1950.
\newblock \doi{10.1103/PhysRev.79.183}.
\newblock URL \url{https://link.aps.org/doi/10.1103/PhysRev.79.183}.

\bibitem[{Alfv{\'e}n}(1942)]{Alfven1942}
H.~{Alfv{\'e}n}.
\newblock {Existence of Electromagnetic-Hydrodynamic Waves}.
\newblock \emph{\nat}, 150:\penalty0 405--406, Oct 1942.
\newblock \doi{10.1038/150405d0}.

\bibitem[{Belcher} and {Davis}(1971)]{Belcher1971}
J.~W. {Belcher} and Jr. {Davis}, Leverett.
\newblock {Large-amplitude Alfv{\'e}n waves in the interplanetary medium, 2}.
\newblock \emph{Journal of Geophysical Research}, 76\penalty0 (16):\penalty0
  3534, Jan 1971.
\newblock \doi{10.1029/JA076i016p03534}.

\bibitem[{Lithwick} et~al.(2007){Lithwick}, {Goldreich}, and
  {Sridhar}]{Lithwick2007}
Y.~{Lithwick}, P.~{Goldreich}, and S.~{Sridhar}.
\newblock {Imbalanced Strong MHD Turbulence}.
\newblock \emph{ApJ}, 655:\penalty0 269--274, Jan 2007.
\newblock \doi{10.1086/509884}.

\bibitem[{Roberts} et~al.(1987{\natexlab{a}}){Roberts}, {Klein}, {Goldstein},
  and {Matthaeus}]{Roberts1987a}
D.~A. {Roberts}, L.~W. {Klein}, M.~L. {Goldstein}, and W.~H. {Matthaeus}.
\newblock {The nature and evolution of magnetohydrodynamic fluctuations in the
  solar wind: Voyager observations}.
\newblock \emph{JGR}, 92\penalty0 (A10):\penalty0 11021--11040, October
  1987{\natexlab{a}}.
\newblock \doi{10.1029/JA092iA10p11021}.

\bibitem[{Roberts} et~al.(1987{\natexlab{b}}){Roberts}, {Goldstein}, {Klein},
  and {Matthaeus}]{Roberts1987b}
D.~A. {Roberts}, M.~L. {Goldstein}, L.~W. {Klein}, and W.~H. {Matthaeus}.
\newblock {Origin and evolution of fluctuations in the solar wind: Helios
  observations and Helios-Voyager comparisons}.
\newblock \emph{JGR}, 92\penalty0 (A11):\penalty0 12023--12035, November
  1987{\natexlab{b}}.
\newblock \doi{10.1029/JA092iA11p12023}.

\bibitem[{Kraichnan}(1965)]{Kraichnan1965}
Robert~H. {Kraichnan}.
\newblock {Inertial-Range Spectrum of Hydromagnetic Turbulence}.
\newblock \emph{Physics of Fluids}, 8:\penalty0 1385--1387, Jul 1965.
\newblock \doi{10.1063/1.1761412}.

\bibitem[{Coleman}(1968)]{Coleman1968}
Jr. {Coleman}, Paul~J.
\newblock {Turbulence, Viscosity, and Dissipation in the Solar-Wind Plasma}.
\newblock \emph{The Astrophysical Journal}, 153:\penalty0 371, Aug 1968.
\newblock \doi{10.1086/149674}.

\bibitem[{Goldreich} and {Sridhar}(1995)]{GS95}
P.~{Goldreich} and S.~{Sridhar}.
\newblock {Toward a Theory of Interstellar Turbulence. II. Strong Alfvenic
  Turbulence}.
\newblock \emph{The Astrophysical Journal}, 438:\penalty0 763, Jan 1995.
\newblock \doi{10.1086/175121}.

\bibitem[{Schekochihin} et~al.(2009){Schekochihin}, {Cowley}, {Dorland},
  {Hammett}, {Howes}, {Quataert}, and {Tatsuno}]{Schekochihin2009}
A.~A. {Schekochihin}, S.~C. {Cowley}, W.~{Dorland}, G.~W. {Hammett}, G.~G.
  {Howes}, E.~{Quataert}, and T.~{Tatsuno}.
\newblock {Astrophysical Gyrokinetics: Kinetic and Fluid Turbulent Cascades in
  Magnetized Weakly Collisional Plasmas}.
\newblock \emph{The Astrophysical Journal Supplement}, 182:\penalty0 310--377,
  May 2009.
\newblock \doi{10.1088/0067-0049/182/1/310}.

\bibitem[{Dobrowolny} et~al.(1980){Dobrowolny}, {Mangeney}, and
  {Veltri}]{Dobrowolny1980}
M.~{Dobrowolny}, A.~{Mangeney}, and P.~{Veltri}.
\newblock {Fully Developed Anisotropic Hydromagnetic Turbulence in
  Interplanetary Space}.
\newblock \emph{Physical Review Letters}, 45\penalty0 (2):\penalty0 144--147,
  Jul 1980.
\newblock \doi{10.1103/PhysRevLett.45.144}.

\bibitem[{Grappin} et~al.(1982){Grappin}, {Frisch}, {Pouquet}, and
  {Leorat}]{Grappin1982}
R.~{Grappin}, U.~{Frisch}, A.~{Pouquet}, and J.~{Leorat}.
\newblock {Alfvenic fluctuations as asymptotic states of MHD turbulence}.
\newblock \emph{Astron \& Astrophys.}, 105\penalty0 (1):\penalty0 6--14,
  January 1982.

\bibitem[{Kolmogorov}(1941)]{Kolmogorov1941}
A.~{Kolmogorov}.
\newblock {The Local Structure of Turbulence in Incompressible Viscous Fluid
  for Very Large Reynolds' Numbers}.
\newblock \emph{Akademiia Nauk SSSR Doklady}, 30:\penalty0 301--305, Jan 1941.

\bibitem[{Boldyrev}(2005)]{Boldyrev2005}
Stanislav {Boldyrev}.
\newblock {On the Spectrum of Magnetohydrodynamic Turbulence}.
\newblock \emph{ApJL}, 626\penalty0 (1):\penalty0 L37--L40, June 2005.
\newblock \doi{10.1086/431649}.

\bibitem[{Boldyrev}(2006)]{Boldyrev2006}
S.~{Boldyrev}.
\newblock {Spectrum of Magnetohydrodynamic Turbulence}.
\newblock \emph{Physical Review Letters}, 96\penalty0 (11):\penalty0 115002,
  March 2006.
\newblock \doi{10.1103/PhysRevLett.96.115002}.

\bibitem[{Mason} et~al.(2006){Mason}, {Cattaneo}, and {Boldyrev}]{Mason2006}
Joanne {Mason}, Fausto {Cattaneo}, and Stanislav {Boldyrev}.
\newblock {Dynamic Alignment in Driven Magnetohydrodynamic Turbulence}.
\newblock \emph{\prl}, 97\penalty0 (25):\penalty0 255002, December 2006.
\newblock \doi{10.1103/PhysRevLett.97.255002}.

\bibitem[{Chandran} et~al.(2015){Chandran}, {Schekochihin}, and
  {Mallet}]{Chandran2015}
B.~D.~G. {Chandran}, A.~A. {Schekochihin}, and A.~{Mallet}.
\newblock {Intermittency and Alignment in Strong RMHD Turbulence}.
\newblock \emph{The Astrophysical Journal}, 807\penalty0 (1):\penalty0 39, Jul
  2015.
\newblock \doi{10.1088/0004-637X/807/1/39}.

\bibitem[{Howes}(2015)]{Howes2015}
Gregory~G. {Howes}.
\newblock {The inherently three-dimensional nature of magnetized plasma
  turbulence}.
\newblock \emph{Journal of Plasma Physics}, 81\penalty0 (2):\penalty0
  325810203, April 2015.
\newblock \doi{10.1017/S0022377814001056}.

\bibitem[{Strauss}(1976)]{Strauss1976}
H.~R. {Strauss}.
\newblock {Nonlinear, three-dimensional magnetohydrodynamics of noncircular
  tokamaks}.
\newblock \emph{Physics of Fluids}, 19\penalty0 (1):\penalty0 134--140, January
  1976.
\newblock \doi{10.1063/1.861310}.

\bibitem[{Beresnyak} and {Lazarian}(2006)]{Beresnyak2006}
A.~{Beresnyak} and A.~{Lazarian}.
\newblock {Polarization Intermittency and Its Influence on MHD Turbulence}.
\newblock \emph{ApJL}, 640\penalty0 (2):\penalty0 L175--L178, April 2006.

\bibitem[{Podesta} et~al.(2009){Podesta}, {Chandran}, {Bhattacharjee},
  {Roberts}, and {Goldstein}]{Podesta2009}
J.~J. {Podesta}, B.~D.~G. {Chandran}, A.~{Bhattacharjee}, D.~A. {Roberts}, and
  M.~L. {Goldstein}.
\newblock {Scale-dependent angle of alignment between velocity and magnetic
  field fluctuations in solar wind turbulence}.
\newblock \emph{JGR}, 114\penalty0 (4):\penalty0 A01107, January 2009.
\newblock \doi{10.1029/2008JA013504}.

\bibitem[{Mallet} et~al.(2015){Mallet}, {Schekochihin}, and
  {Chandran}]{Mallet2015}
A.~{Mallet}, A.~A. {Schekochihin}, and B.~D.~G. {Chandran}.
\newblock {Refined critical balance in strong Alfvenic turbulence.}
\newblock \emph{MNRAS}, 449:\penalty0 L77--L81, April 2015.

\bibitem[{Mallet} et~al.(2016)]{Mallet2016}
A.~{Mallet} et~al.
\newblock {Measures of three-dimensional anisotropy and intermittency in strong
  Alfv{\'e}nic turbulence}.
\newblock \emph{MNRAS}, 459\penalty0 (2):\penalty0 2130--2139, June 2016.

\bibitem[{Goldstein} et~al.(1974){Goldstein}, {Klimas}, and
  {Barish}]{Goldstein1974}
M.~L. {Goldstein}, A.~J. {Klimas}, and F.~D. {Barish}.
\newblock {On the theory of large amplitude Alfv{\'e}n waves.}
\newblock In C.~T. {Russell}, editor, \emph{Solar Wind Three}, pages 385--387,
  January 1974.

\bibitem[{Lichtenstein} and {Sonett}(1980)]{Lichtenstein1980}
B.~R. {Lichtenstein} and C.~P. {Sonett}.
\newblock {Dynamic magnetic structure of large amplitude Alfv{\'e}nic
  variations in the solar wind}.
\newblock \emph{GRL}, 7\penalty0 (3):\penalty0 189--192, March 1980.
\newblock \doi{10.1029/GL007i003p00189}.

\bibitem[{Riley} et~al.(1996){Riley}, {Sonett}, {Tsurutani}, {Balogh},
  {Forsyth}, and {Hoogeveen}]{Riley1996}
Pete {Riley}, C.~P. {Sonett}, B.~T. {Tsurutani}, A.~{Balogh}, R.~J. {Forsyth},
  and G.~W. {Hoogeveen}.
\newblock {Properties of arc-polarized Alfv{\'e}n waves in the ecliptic plane:
  Ulysses observations}.
\newblock \emph{JGR}, 101\penalty0 (A9):\penalty0 19987--19994, September 1996.
\newblock \doi{10.1029/96JA01743}.

\bibitem[{Barnes} and {Hollweg}(1974)]{BarnesHollweg1974}
Aaron {Barnes} and Joseph~V. {Hollweg}.
\newblock {Large-amplitude hydromagnetic waves}.
\newblock \emph{JGR}, 79\penalty0 (16):\penalty0 2302, January 1974.
\newblock \doi{10.1029/JA079i016p02302}.

\bibitem[Bale et~al.(2019)Bale, Badman, Bonnell, Bowen, Burgess, Case, Cattell,
  Chandran, Chaston, Chen, et~al.]{Bale2019}
SD~Bale, ST~Badman, JW~Bonnell, TA~Bowen, D~Burgess, AW~Case, CA~Cattell, BDG
  Chandran, CC~Chaston, CHK Chen, et~al.
\newblock Highly structured slow solar wind emerging from an equatorial coronal
  hole.
\newblock \emph{Nature}, pages 1--6, 2019.

\bibitem[{Dudok de Wit} et~al.(2020){Dudok de Wit}, {Krasnoselskikh}, {Bale},
  {Bonnell}, {Bowen}, {Chen}, {Froment}, {Goetz}, {Harvey}, {Jagarlamudi},
  {Larosa}, {MacDowall}, {Malaspina}, {Matthaeus}, {Pulupa}, {Velli}, and
  {Whittlesey}]{DudokdeWit2020}
Thierry {Dudok de Wit}, Vladimir~V. {Krasnoselskikh}, Stuart~D. {Bale}, John~W.
  {Bonnell}, Trevor~A. {Bowen}, Christopher H.~K. {Chen}, Clara {Froment},
  Keith {Goetz}, Peter~R. {Harvey}, Vamsee~Krishna {Jagarlamudi}, Andrea
  {Larosa}, Robert~J. {MacDowall}, David~M. {Malaspina}, William~H.
  {Matthaeus}, Marc {Pulupa}, Marco {Velli}, and Phyllis~L. {Whittlesey}.
\newblock {Switchbacks in the Near-Sun Magnetic Field: Long Memory and Impact
  on the Turbulence Cascade}.
\newblock \emph{ApJS}, 246\penalty0 (2):\penalty0 39, February 2020.
\newblock \doi{10.3847/1538-4365/ab5853}.

\bibitem[{Horbury} et~al.(2020){Horbury}, {Woolley}, {Laker}, {Matteini},
  {Eastwood}, {Bale}, {Velli}, {Chandran}, {Phan}, {Raouafi}, {Goetz},
  {Harvey}, {Pulupa}, {Klein}, {Dudok de Wit}, {Kasper}, {Korreck}, {Case},
  {Stevens}, {Whittlesey}, {Larson}, {MacDowall}, {Malaspina}, and
  {Livi}]{Horbury2020}
Timothy~S. {Horbury}, Thomas {Woolley}, Ronan {Laker}, Lorenzo {Matteini},
  Jonathan {Eastwood}, Stuart~D. {Bale}, Marco {Velli}, Benjamin D.~G.
  {Chandran}, Tai {Phan}, Nour~E. {Raouafi}, Keith {Goetz}, Peter~R. {Harvey},
  Marc {Pulupa}, K.~G. {Klein}, Thierry {Dudok de Wit}, Justin~C. {Kasper},
  Kelly~E. {Korreck}, A.~W. {Case}, Michael~L. {Stevens}, Phyllis {Whittlesey},
  Davin {Larson}, Robert~J. {MacDowall}, David~M. {Malaspina}, and Roberto
  {Livi}.
\newblock {Sharp Alfv{\'e}nic Impulses in the Near-Sun Solar Wind}.
\newblock \emph{ApJS}, 246\penalty0 (2):\penalty0 45, February 2020.
\newblock \doi{10.3847/1538-4365/ab5b15}.

\bibitem[{Chen} et~al.(2020){Chen}, {Bale}, {Bonnell}, {Borovikov}, {Bowen},
  {Burgess}, {Case}, {Chandran}, {de Wit}, {Goetz}, {Harvey}, {Kasper},
  {Klein}, {Korreck}, {Larson}, {Livi}, {MacDowall}, {Malaspina}, {Mallet},
  {McManus}, {Moncuquet}, {Pulupa}, {Stevens}, and {Whittlesey}]{Chen2020}
C.~H.~K. {Chen}, S.~D. {Bale}, J.~W. {Bonnell}, D.~{Borovikov}, T.~A. {Bowen},
  D.~{Burgess}, A.~W. {Case}, B.~D.~G. {Chandran}, T.~Dudok {de Wit},
  K.~{Goetz}, P.~R. {Harvey}, J.~C. {Kasper}, K.~G. {Klein}, K.~E. {Korreck},
  D.~{Larson}, R.~{Livi}, R.~J. {MacDowall}, D.~M. {Malaspina}, A.~{Mallet},
  M.~D. {McManus}, M.~{Moncuquet}, M.~{Pulupa}, M.~L. {Stevens}, and
  P.~{Whittlesey}.
\newblock {The Evolution and Role of Solar Wind Turbulence in the Inner
  Heliosphere}.
\newblock \emph{ApJS}, 246\penalty0 (2):\penalty0 53, February 2020.
\newblock \doi{10.3847/1538-4365/ab60a3}.

\bibitem[{McManus} et~al.(2020){McManus}, {Bowen}, {Mallet}, {Chen},
  {Chandran}, {Bale}, {Larson}, {Dudok de Wit}, {Kasper}, {Stevens},
  {Whittlesey}, {Livi}, {Korreck}, {Goetz}, {Harvey}, {Pulupa}, {MacDowall},
  {Malaspina}, {Case}, and {Bonnell}]{McManus2020}
Michael~D. {McManus}, Trevor~A. {Bowen}, Alfred {Mallet}, Christopher H.~K.
  {Chen}, Benjamin D.~G. {Chandran}, Stuart~D. {Bale}, Davin~E. {Larson},
  Thierry {Dudok de Wit}, J.~C. {Kasper}, Michael {Stevens}, Phyllis
  {Whittlesey}, Roberto {Livi}, Kelly~E. {Korreck}, Keith {Goetz}, Peter~R.
  {Harvey}, Marc {Pulupa}, Robert~J. {MacDowall}, David~M. {Malaspina},
  Anthony~W. {Case}, and J.~W. {Bonnell}.
\newblock {Cross Helicity Reversals in Magnetic Switchbacks}.
\newblock \emph{ApJS}, 246\penalty0 (2):\penalty0 67, February 2020.
\newblock \doi{10.3847/1538-4365/ab6dce}.

\bibitem[{Chaston} et~al.(2020){Chaston}, {Bonnell}, {Bale}, {Kasper},
  {Pulupa}, {Dudok de Wit}, {{{Bowen}}}, et~al.]{Chaston2020}
C.~C. {Chaston}, J.~W. {Bonnell}, S.~D. {Bale}, J.~C. {Kasper}, M.~{Pulupa},
  T.~{Dudok de Wit}, {{{T.~A.}}} {{{Bowen}}}, et~al.
\newblock {MHD Mode Composition in the Inner Heliosphere from the Parker Solar
  Probe{\textquoteright}s First Perihelion}.
\newblock \emph{ApJS}, 246\penalty0 (2):\penalty0 71, February 2020.
\newblock \doi{10.3847/1538-4365/ab745c}.
\newblock URL \url{https://ui.adsabs.harvard.edu/abs/2020ApJS..246...71C}.

\bibitem[{Bourouaine} et~al.(2020){Bourouaine}, {Perez}, {Klein}, {Chen},
  {Martinovi{\'c}}, {Bale}, {Kasper}, and {Raouafi}]{Bourouaine2020}
Sofiane {Bourouaine}, Jean~C. {Perez}, Kristopher~G. {Klein}, Christopher H.~K.
  {Chen}, Mihailo {Martinovi{\'c}}, Stuart~D. {Bale}, Justin~C. {Kasper}, and
  Nour~E. {Raouafi}.
\newblock {Turbulence Characteristics of Switchback and Nonswitchback Intervals
  Observed by Parker Solar Probe}.
\newblock \emph{ApJL}, 904\penalty0 (2):\penalty0 L30, December 2020.
\newblock \doi{10.3847/2041-8213/abbd4a}.

\bibitem[{Chen} et~al.(2021)]{Chen2021}
C.~H.~K. {Chen} et~al.
\newblock {The Near-Sun Streamer Belt Solar Wind: Turbulence and Solar Wind
  Acceleration}.
\newblock \emph{A\&A}, January 2021.

\bibitem[{Martinovi{\'c}} et~al.(2021){Martinovi{\'c}}, {Klein}, {Huang},
  {Chandran}, {Kasper}, {Lichko}, {{{Bowen}}}, et~al.]{Martinovic2021}
M.~M. {Martinovi{\'c}}, K.~G. {Klein}, J.~{Huang}, B.~D.~G. {Chandran}, J.~C.
  {Kasper}, E.~{Lichko}, {{{T.~A.}}} {{{Bowen}}}, et~al.
\newblock {Multiscale Solar Wind Turbulence Properties inside and near
  Switchbacks measured by Parker Solar Probe}.
\newblock \emph{arXiv e-prints; Accepted/ApJ}, art. arXiv:2103.00374, February
  2021.

\bibitem[{Wang} et~al.(2012)]{Wang2012}
Xin {Wang} et~al.
\newblock {Large-amplitude Alfv{\'e}n Wave in Interplanetary Space: The Wind
  Spacecraft Observations}.
\newblock \emph{ApJ}, 746\penalty0 (2):\penalty0 147, February 2012.

\bibitem[{Matteini} et~al.(2015){Matteini}, {Horbury}, {Pantellini}, {Velli},
  and {Schwartz}]{Matteini2015}
L.~{Matteini}, T.~S. {Horbury}, F.~{Pantellini}, M.~{Velli}, and S.~J.
  {Schwartz}.
\newblock {Ion Kinetic Energy Conservation and Magnetic Field Strength
  Constancy in Multi-fluid Solar Wind Alfv{\'e}nic Turbulence}.
\newblock \emph{ApJ}, 802\penalty0 (1):\penalty0 11, March 2015.
\newblock \doi{10.1088/0004-637X/802/1/11}.

\bibitem[{de Hoffmann} and {Teller}(1950)]{deHoffmannTeller1950}
F.~{de Hoffmann} and E.~{Teller}.
\newblock {Magneto-Hydrodynamic Shocks}.
\newblock \emph{Physical Review}, 80\penalty0 (4):\penalty0 692--703, November
  1950.
\newblock \doi{10.1103/PhysRev.80.692}.

\bibitem[{Hollweg}(1974)]{Hollweg1974}
Joseph~V. {Hollweg}.
\newblock {Transverse Alfv{\'e}n waves in the solar wind: Arbitrary k, v
  $_{0}$, B $_{0}$, and |{\ensuremath{\delta}}B|}.
\newblock \emph{JGR}, 79\penalty0 (10):\penalty0 1539, January 1974.
\newblock \doi{10.1029/JA079i010p01539}.

\bibitem[{Chen} et~al.(2013){Chen}, {Bale}, {Salem}, and {Maruca}]{Chen2013}
C.~H.~K. {Chen}, S.~D. {Bale}, C.~S. {Salem}, and B.~A. {Maruca}.
\newblock {Residual Energy Spectrum of Solar Wind Turbulence}.
\newblock \emph{ApJ}, 770\penalty0 (2):\penalty0 125, June 2013.
\newblock \doi{10.1088/0004-637X/770/2/125}.

\bibitem[{Bowen} et~al.(2018{\natexlab{a}}){Bowen}, {Mallet}, {Bonnell}, and
  {Bale}]{Bowen2018b}
T.~A. {Bowen}, A.~{Mallet}, J.~W. {Bonnell}, and S.~D. {Bale}.
\newblock {Impact of Residual Energy on Solar Wind Turbulent Spectra}.
\newblock \emph{The Astrophysical Journal}, 865:\penalty0 45, Sep
  2018{\natexlab{a}}.
\newblock \doi{10.3847/1538-4357/aad95b}.

\bibitem[{Velli} et~al.(1989){Velli}, {Grappin}, and {Mangeney}]{Velli1989}
Marco {Velli}, Roland {Grappin}, and Andre {Mangeney}.
\newblock {Turbulent cascade of incompressible unidirectional Alfv{\'e}n waves
  in the interplanetary medium}.
\newblock \emph{\prl}, 63\penalty0 (17):\penalty0 1807--1810, October 1989.
\newblock \doi{10.1103/PhysRevLett.63.1807}.

\bibitem[{Perez} and {Chandran}(2013)]{Perez2013}
Jean~Carlos {Perez} and Benjamin D.~G. {Chandran}.
\newblock {Direct Numerical Simulations of Reflection-driven, Reduced
  Magnetohydrodynamic Turbulence from the Sun to the Alfv{\'e}n Critical
  Point}.
\newblock \emph{\apj}, 776\penalty0 (2):\penalty0 124, October 2013.
\newblock \doi{10.1088/0004-637X/776/2/124}.

\bibitem[{Chandran} and {Perez}(2019)]{Chandran2019}
Benjamin D.~G. {Chandran} and Jean~C. {Perez}.
\newblock {Reflection-driven magnetohydrodynamic turbulence in the solar
  atmosphere and solar wind}.
\newblock \emph{Journal of Plasma Physics}, 85\penalty0 (4):\penalty0
  905850409, August 2019.
\newblock \doi{10.1017/S0022377819000540}.

\bibitem[{Bale} et~al.(2016){Bale}, {Goetz}, {Harvey}, {Turin}, {Bonnell},
  {Dudok de Wit}, {Ergun}, {MacDowall}, {Pulupa}, {Andre}, {Bolton},
  {Bougeret}, {Bowen}, {Burgess}, {Cattell}, {Chandran}, {Chaston}, {Chen},
  {Choi}, {Connerney}, {Cranmer}, {Diaz-Aguado}, {Donakowski}, {Drake},
  {Farrell}, {Fergeau}, {Fermin}, {Fischer}, {Fox}, {Glaser}, {Goldstein},
  {Gordon}, {Hanson}, {Harris}, {Hayes}, {Hinze}, {Hollweg}, {Horbury},
  {Howard}, {Hoxie}, {Jannet}, {Karlsson}, {Kasper}, {Kellogg}, {Kien},
  {Klimchuk}, {Krasnoselskikh}, {Krucker}, {Lynch}, {Maksimovic}, {Malaspina},
  {Marker}, {Martin}, {Martinez-Oliveros}, {McCauley}, {McComas}, {McDonald},
  {Meyer-Vernet}, {Moncuquet}, {Monson}, {Mozer}, {Murphy}, {Odom},
  {Oliverson}, {Olson}, {Parker}, {Pankow}, {Phan}, {Quataert}, {Quinn},
  {Ruplin}, {Salem}, {Seitz}, {Sheppard}, {Siy}, {Stevens}, {Summers}, {Szabo},
  {Timofeeva}, {Vaivads}, {Velli}, {Yehle}, {Werthimer}, and
  {Wygant}]{Bale2016}
S.~D. {Bale}, K.~{Goetz}, P.~R. {Harvey}, P.~{Turin}, J.~W. {Bonnell},
  T.~{Dudok de Wit}, R.~E. {Ergun}, R.~J. {MacDowall}, M.~{Pulupa}, M.~{Andre},
  M.~{Bolton}, J.-L. {Bougeret}, T.~A. {Bowen}, D.~{Burgess}, C.~A. {Cattell},
  B.~D.~G. {Chandran}, C.~C. {Chaston}, C.~H.~K. {Chen}, M.~K. {Choi}, J.~E.
  {Connerney}, S.~{Cranmer}, M.~{Diaz-Aguado}, W.~{Donakowski}, J.~F. {Drake},
  W.~M. {Farrell}, P.~{Fergeau}, J.~{Fermin}, J.~{Fischer}, N.~{Fox},
  D.~{Glaser}, M.~{Goldstein}, D.~{Gordon}, E.~{Hanson}, S.~E. {Harris}, L.~M.
  {Hayes}, J.~J. {Hinze}, J.~V. {Hollweg}, T.~S. {Horbury}, R.~A. {Howard},
  V.~{Hoxie}, G.~{Jannet}, M.~{Karlsson}, J.~C. {Kasper}, P.~J. {Kellogg},
  M.~{Kien}, J.~A. {Klimchuk}, V.~V. {Krasnoselskikh}, S.~{Krucker}, J.~J.
  {Lynch}, M.~{Maksimovic}, D.~M. {Malaspina}, S.~{Marker}, P.~{Martin},
  J.~{Martinez-Oliveros}, J.~{McCauley}, D.~J. {McComas}, T.~{McDonald},
  N.~{Meyer-Vernet}, M.~{Moncuquet}, S.~J. {Monson}, F.~S. {Mozer}, S.~D.
  {Murphy}, J.~{Odom}, R.~{Oliverson}, J.~{Olson}, E.~N. {Parker}, D.~{Pankow},
  T.~{Phan}, E.~{Quataert}, T.~{Quinn}, S.~W. {Ruplin}, C.~{Salem}, D.~{Seitz},
  D.~A. {Sheppard}, A.~{Siy}, K.~{Stevens}, D.~{Summers}, A.~{Szabo},
  M.~{Timofeeva}, A.~{Vaivads}, M.~{Velli}, A.~{Yehle}, D.~{Werthimer}, and
  J.~R. {Wygant}.
\newblock {The FIELDS Instrument Suite for Solar Probe Plus. Measuring the
  Coronal Plasma and Magnetic Field, Plasma Waves and Turbulence, and Radio
  Signatures of Solar Transients}.
\newblock \emph{Space Science Rev.}, 204:\penalty0 49--82, December 2016.
\newblock \doi{10.1007/s11214-016-0244-5}.

\bibitem[Kasper et~al.(2016)Kasper, Abiad, Austin, Balat-Pichelin, Bale,
  Belcher, Berg, Bergner, Berthomier, Bookbinder, Brodu, Caldwell, Case,
  Chandran, Cheimets, Cirtain, Cranmer, Curtis, Daigneau, Dalton, Dasgupta,
  DeTomaso, Diaz-Aguado, Djordjevic, Donaskowski, Effinger, Florinski, Fox,
  Freeman, Gallagher, Gary, Gauron, Gates, Goldstein, Golub, Gordon, Gurnee,
  Guth, Halekas, Hatch, Heerikuisen, Ho, Hu, Johnson, Jordan, Korreck, Larson,
  Lazarus, Li, Livi, Ludlam, Maksimovic, McFadden, Marchant, Maruca, McComas,
  Messina, Mercer, Park, Peddie, Pogorelov, Reinhart, Richardson, Robinson,
  Rosen, Skoug, Slagle, Steinberg, Stevens, Szabo, Taylor, Tiu, Turin, Velli,
  Webb, Whittlesey, Wright, Wu, and Zank]{Kasper2016}
Justin~C. Kasper, Robert Abiad, Gerry Austin, Marianne Balat-Pichelin,
  Stuart~D. Bale, John~W. Belcher, Peter Berg, Henry Bergner, Matthieu
  Berthomier, Jay Bookbinder, Etienne Brodu, David Caldwell, Anthony~W. Case,
  Benjamin D.~G. Chandran, Peter Cheimets, Jonathan~W. Cirtain, Steven~R.
  Cranmer, David~W. Curtis, Peter Daigneau, Greg Dalton, Brahmananda Dasgupta,
  David DeTomaso, Millan Diaz-Aguado, Blagoje Djordjevic, Bill Donaskowski,
  Michael Effinger, Vladimir Florinski, Nichola Fox, Mark Freeman, Dennis
  Gallagher, S.~Peter Gary, Tom Gauron, Richard Gates, Melvin Goldstein, Leon
  Golub, Dorothy~A. Gordon, Reid Gurnee, Giora Guth, Jasper Halekas, Ken Hatch,
  Jacob Heerikuisen, George Ho, Qiang Hu, Greg Johnson, Steven~P. Jordan,
  Kelly~E. Korreck, Davin Larson, Alan~J. Lazarus, Gang Li, Roberto Livi,
  Michael Ludlam, Milan Maksimovic, James~P. McFadden, William Marchant,
  Bennet~A. Maruca, David~J. McComas, Luciana Messina, Tony Mercer, Sang Park,
  Andrew~M. Peddie, Nikolai Pogorelov, Matthew~J. Reinhart, John~D. Richardson,
  Miles Robinson, Irene Rosen, Ruth~M. Skoug, Amanda Slagle, John~T. Steinberg,
  Michael~L. Stevens, Adam Szabo, Ellen~R. Taylor, Chris Tiu, Paul Turin, Marco
  Velli, Gary Webb, Phyllis Whittlesey, Ken Wright, S.~T. Wu, and Gary Zank.
\newblock Solar wind electrons alphas and protons (sweap) investigation: Design
  of the solar wind and coronal plasma instrument suite for solar probe plus.
\newblock \emph{Space Science Reviews}, 204\penalty0 (1):\penalty0 131--186,
  Dec 2016.
\newblock ISSN 1572-9672.
\newblock \doi{10.1007/s11214-015-0206-3}.
\newblock URL \url{https://doi.org/10.1007/s11214-015-0206-3}.

\bibitem[{Bowen} et~al.(2020)]{Bowen2020b}
{{{T.~A.}}} {Bowen} et~al.
\newblock A merged search-coil and fluxgate magnetometer data product for
  parker solar probe fields.
\newblock \emph{JGR}, 125\penalty0 (5):\penalty0 e2020JA027813, 2020.

\bibitem[{Klein} et~al.(2021){Klein}, {Verniero}, {Alterman}, {Bale}, {Case},
  {Kasper}, {Korreck}, {Larson}, {Lichko}, {Livi}, {McManus}, {Martinovi{\'c}},
  {Rahmati}, {Stevens}, and {Whittlesey}]{Klein2021}
K.~G. {Klein}, J.~L. {Verniero}, B.~{Alterman}, S.~{Bale}, A.~{Case}, J.~C.
  {Kasper}, K.~{Korreck}, D.~{Larson}, E.~{Lichko}, R.~{Livi}, M.~{McManus},
  M.~{Martinovi{\'c}}, A.~{Rahmati}, M.~{Stevens}, and P.~{Whittlesey}.
\newblock {Inferred Linear Stability of Parker Solar Probe Observations Using
  One- and Two-component Proton Distributions}.
\newblock \emph{ApJ}, 909\penalty0 (1):\penalty0 7, March 2021.
\newblock \doi{10.3847/1538-4357/abd7a0}.

\bibitem[{Malaspina} et~al.(2016){Malaspina}, {Ergun}, {Bolton}, {Kien},
  {Summers}, {Stevens}, {Yehle}, {Karlsson}, {Hoxie}, {Bale}, and
  {Goetz}]{Malaspina2016}
David~M. {Malaspina}, Robert~E. {Ergun}, Mary {Bolton}, Mark {Kien}, David
  {Summers}, Ken {Stevens}, Alan {Yehle}, Magnus {Karlsson}, Vaughn~C. {Hoxie},
  Stuart~D. {Bale}, and Keith {Goetz}.
\newblock {The Digital Fields Board for the FIELDS instrument suite on the
  Solar Probe Plus mission: Analog and digital signal processing}.
\newblock \emph{Journal of Geophysical Research (Space Physics)}, 121:\penalty0
  5088--5096, June 2016.
\newblock \doi{10.1002/2016JA022344}.

\bibitem[{Pulupa} et~al.(2017){Pulupa}, {Bale}, {Bonnell}, {Bowen}, {Carruth},
  {Goetz}, {Gordon}, {Harvey}, {Maksimovic}, {Mart{\'\i}nez-Oliveros},
  {Moncuquet}, {Saint-Hilaire}, {Seitz}, and {Sundkvist}]{Pulupa2017}
M.~{Pulupa}, S.~D. {Bale}, J.~W. {Bonnell}, T.~A. {Bowen}, N.~{Carruth},
  K.~{Goetz}, D.~{Gordon}, P.~R. {Harvey}, M.~{Maksimovic}, J.~C.
  {Mart{\'\i}nez-Oliveros}, M.~{Moncuquet}, P.~{Saint-Hilaire}, D.~{Seitz}, and
  D.~{Sundkvist}.
\newblock {The Solar Probe Plus Radio Frequency Spectrometer: Measurement
  requirements, analog design, and digital signal processing}.
\newblock \emph{Journal of Geophysical Research (Space Physics)}, 122\penalty0
  (3):\penalty0 2836--2854, March 2017.
\newblock \doi{10.1002/2016JA023345}.

\bibitem[{Barnes}(1979)]{Barnes1979}
A.~{Barnes}.
\newblock \emph{{Hydromagnetic waves and turbulence in the solar wind}},
  volume~1, pages 249--319.
\newblock 1979.

\bibitem[{Khrabrov} and {Sonnerup}(1998)]{1998ISSI}
Alexander~V. {Khrabrov} and Bengt U.~{\"O}. {Sonnerup}.
\newblock {DeHoffmann-Teller Analysis}.
\newblock \emph{ISSI Scientific Reports Series}, 1:\penalty0 221--248, January
  1998.

\bibitem[{Wicks} et~al.(2013{\natexlab{a}}){Wicks}, {Mallet}, {Horbury},
  {Chen}, {Schekochihin}, and {Mitchell}]{Wicks2013a}
R.~T. {Wicks}, A.~{Mallet}, T.~S. {Horbury}, C.~H.~K. {Chen}, A.~A.
  {Schekochihin}, and J.~J. {Mitchell}.
\newblock {Alignment and Scaling of Large-Scale Fluctuations in the Solar
  Wind}.
\newblock \emph{PRL}, 110\penalty0 (2):\penalty0 025003, January
  2013{\natexlab{a}}.
\newblock \doi{10.1103/PhysRevLett.110.025003}.

\bibitem[{Wicks} et~al.(2013{\natexlab{b}}){Wicks}, {Roberts}, {Mallet},
  {Schekochihin}, {Horbury}, and {Chen}]{Wicks2013b}
R.~T. {Wicks}, D.~A. {Roberts}, A.~{Mallet}, A.~A. {Schekochihin}, T.~S.
  {Horbury}, and C.~H.~K. {Chen}.
\newblock {Correlations at Large Scales and the Onset of Turbulence in the Fast
  Solar Wind}.
\newblock \emph{ApJ}, 778\penalty0 (2):\penalty0 177, December
  2013{\natexlab{b}}.
\newblock \doi{10.1088/0004-637X/778/2/177}.

\bibitem[{Frisch}(1995)]{Frisch1995}
Uriel {Frisch}.
\newblock \emph{{Turbulence}}.
\newblock 1995.

\bibitem[{Matteini} et~al.(2018){Matteini}, {Stansby}, {Horbury}, and
  {Chen}]{Matteini2018}
L.~{Matteini}, D.~{Stansby}, T.~S. {Horbury}, and C.~H.~K. {Chen}.
\newblock {On the 1/f Spectrum in the Solar Wind and Its Connection with
  Magnetic Compressibility}.
\newblock \emph{ApJL}, 869\penalty0 (2):\penalty0 L32, December 2018.
\newblock \doi{10.3847/2041-8213/aaf573}.

\bibitem[{Matteini} et~al.(2019){Matteini}, {Stansby}, {Horbury}, and
  {Chen}]{Matteini2019}
L.~{Matteini}, D.~{Stansby}, T.~S. {Horbury}, and C.~H.~K. {Chen}.
\newblock {The rotation angle distribution underlying magnetic field
  fluctuations in the 1/ f range of solar wind turbulent spectra}.
\newblock \emph{Nuovo Cimento C Geophysics Space Physics C}, 42\penalty0
  (1):\penalty0 16, January 2019.
\newblock \doi{10.1393/ncc/i2019-19016-y}.

\bibitem[{Cohen} and {Kulsrud}(1974)]{CohenKulsrud1974}
R.~H. {Cohen} and R.~M. {Kulsrud}.
\newblock {Nonlinear evolution of parallel-propagating hydromagnetic waves.}
\newblock \emph{Physics of Fluids}, 17:\penalty0 2215--2225, December 1974.
\newblock \doi{10.1063/1.1694695}.

\bibitem[{Cranmer} and {van Ballegooijen}(2005)]{Cranmer2005}
S.~R. {Cranmer} and A.~A. {van Ballegooijen}.
\newblock {On the Generation, Propagation, and Reflection of Alfv{\'e}n Waves
  from the Solar Photosphere to the Distant Heliosphere}.
\newblock \emph{ApJS}, 156\penalty0 (2):\penalty0 265--293, February 2005.
\newblock \doi{10.1086/426507}.

\bibitem[{Verdini} and {Velli}(2007)]{Verdini2007}
Andrea {Verdini} and Marco {Velli}.
\newblock {Alfv{\'e}n Waves and Turbulence in the Solar Atmosphere and Solar
  Wind}.
\newblock \emph{ApJ}, 662\penalty0 (1):\penalty0 669--676, June 2007.
\newblock \doi{10.1086/510710}.

\bibitem[{Squire} et~al.(2020){Squire}, {Chandran}, and {Meyrand}]{Squire2020}
J.~{Squire}, B.~D.~G. {Chandran}, and R.~{Meyrand}.
\newblock {In-situ Switchback Formation in the Expanding Solar Wind}.
\newblock \emph{ApJL}, 891\penalty0 (1):\penalty0 L2, March 2020.
\newblock \doi{10.3847/2041-8213/ab74e1}.

\bibitem[{Shoda} et~al.(2021){Shoda}, {Chandran}, and {Cranmer}]{Shoda2021}
Munehito {Shoda}, Benjamin D.~G. {Chandran}, and Steven~R. {Cranmer}.
\newblock {Turbulent generation of magnetic switchbacks in the Alfv{\'e}nic
  solar wind}.
\newblock \emph{arXiv e-prints}, art. arXiv:2101.09529, January 2021.

\bibitem[{Mallet} et~al.(2021){Mallet}, {Squire}, {Chandran}, {Bowen}, and
  {Bale}]{Mallet2021}
Alfred {Mallet}, Jonathan {Squire}, Benjamin D.~G. {Chandran}, Trevor {Bowen},
  and Stuart~D. {Bale}.
\newblock {Evolution of large-amplitude Alfv{\'e}n waves and generation of
  switchbacks in the expanding solar wind}.
\newblock \emph{arXiv e-prints}, art. arXiv:2104.08321, April 2021.

\bibitem[{Servidio} et~al.(2008){Servidio}, {Matthaeus}, and
  {Dmitruk}]{Servidio2008}
S.~{Servidio}, W.~H. {Matthaeus}, and P.~{Dmitruk}.
\newblock {Depression of Nonlinearity in Decaying Isotropic MHD Turbulence}.
\newblock \emph{\prl}, 100\penalty0 (9):\penalty0 095005, March 2008.
\newblock \doi{10.1103/PhysRevLett.100.095005}.

\bibitem[{Derby}(1978)]{Derby1978}
Jr. {Derby}, N.~F.
\newblock {Modulational instability of finite-amplitude, circularly polarized
  Alfv{\'e}n waves.}
\newblock \emph{ApJ}, 224:\penalty0 1013--1016, September 1978.
\newblock \doi{10.1086/156451}.

\bibitem[{Vi{\~n}as} and {Goldstein}(1991)]{Vinas1991}
Adolfo~F. {Vi{\~n}as} and Melvyn~L. {Goldstein}.
\newblock {Parametric instabilities of circularly polarized large-amplitude
  dispersive Alfv{\'e}n waves: excitation of obliquely-propagating daughter and
  side-band waves}.
\newblock \emph{Journal of Plasma Physics}, 46\penalty0 (1):\penalty0 129--152,
  August 1991.
\newblock \doi{10.1017/S0022377800015993}.

\bibitem[{Del Zanna}(2001)]{DelZanna2001}
Luca {Del Zanna}.
\newblock {Parametric decay of oblique arc-polarized Alfv{\'e}n waves}.
\newblock \emph{GRL}, 28\penalty0 (13):\penalty0 2585--2588, January 2001.
\newblock \doi{10.1029/2001GL012911}.

\bibitem[{Tenerani} and {Velli}(2013)]{Tenerani2013}
A.~{Tenerani} and M.~{Velli}.
\newblock {Parametric decay of radial Alfv{\'e}n waves in the expanding
  accelerating solar wind}.
\newblock \emph{Journal of Geophysical Research (Space Physics)}, 118\penalty0
  (12):\penalty0 7507--7516, December 2013.
\newblock \doi{10.1002/2013JA019293}.

\bibitem[{Squire} et~al.(2017){Squire}, {Kunz}, {Quataert}, and
  {Schekochihin}]{Squire2017}
J.~{Squire}, M.~W. {Kunz}, E.~{Quataert}, and A.~A. {Schekochihin}.
\newblock {Kinetic Simulations of the Interruption of Large-Amplitude
  Shear-Alfv{\'e}n Waves in a High-{\ensuremath{\beta}} Plasma}.
\newblock \emph{PRL}, 119\penalty0 (15):\penalty0 155101, October 2017.
\newblock \doi{10.1103/PhysRevLett.119.155101}.

\bibitem[{Bowen} et~al.(2018{\natexlab{b}}){Bowen}, {Badman}, {Hellinger}, and
  {Bale}]{Bowen2018a}
Trevor~A. {Bowen}, Samuel {Badman}, Petr {Hellinger}, and Stuart~D. {Bale}.
\newblock {Density Fluctuations in the Solar Wind Driven by Alfv{\'e}n Wave
  Parametric Decay}.
\newblock \emph{ApJL}, 854\penalty0 (2):\penalty0 L33, February
  2018{\natexlab{b}}.
\newblock \doi{10.3847/2041-8213/aaabbe}.

\bibitem[{Chandran}(2018)]{Chandran2018}
Benjamin D.~G. {Chandran}.
\newblock {Parametric instability, inverse cascade and the 1/f range of
  solar-wind turbulence}.
\newblock \emph{Journal of Plasma Physics}, 84\penalty0 (1):\penalty0
  905840106, February 2018.
\newblock \doi{10.1017/S0022377818000016}.

\bibitem[{Gonz{\'a}lez} et~al.(2020){Gonz{\'a}lez}, {Tenerani}, {Velli}, and
  {Hellinger}]{Gonzalez2020}
C.~A. {Gonz{\'a}lez}, A.~{Tenerani}, M.~{Velli}, and P.~{Hellinger}.
\newblock {The Role of Parametric Instabilities in Turbulence Generation and
  Proton Heating: Hybrid Simulations of Parallel-propagating Alfv{\'e}n Waves}.
\newblock \emph{ApJ}, 904\penalty0 (1):\penalty0 81, November 2020.
\newblock \doi{10.3847/1538-4357/abbccd}.

\bibitem[{Tenerani} et~al.(2020){Tenerani}, {Velli}, {Matteini}, {R{\'e}ville},
  {Shi}, {Bale}, {Kasper}, {Bonnell}, {Case}, {de Wit}, {Goetz}, {Harvey},
  {Klein}, {Korreck}, {Larson}, {Livi}, {MacDowall}, {Malaspina}, {Pulupa},
  {Stevens}, and {Whittlesey}]{Tenerani2020}
Anna {Tenerani}, Marco {Velli}, Lorenzo {Matteini}, Victor {R{\'e}ville}, Chen
  {Shi}, Stuart~D. {Bale}, Justin~C. {Kasper}, John~W. {Bonnell}, Anthony~W.
  {Case}, Thierry~Dudok {de Wit}, Keith {Goetz}, Peter~R. {Harvey},
  Kristopher~G. {Klein}, Kelly {Korreck}, Davin {Larson}, Roberto {Livi},
  Robert~J. {MacDowall}, David~M. {Malaspina}, Marc {Pulupa}, Michael
  {Stevens}, and Phyllis {Whittlesey}.
\newblock {Magnetic Field Kinks and Folds in the Solar Wind}.
\newblock \emph{ApJS}, 246\penalty0 (2):\penalty0 32, February 2020.
\newblock \doi{10.3847/1538-4365/ab53e1}.

\bibitem[{Matteini} et~al.(2010){Matteini}, {Landi}, {Velli}, and
  {Hellinger}]{Matteini2010}
Lorenzo {Matteini}, Simone {Landi}, Marco {Velli}, and Petr {Hellinger}.
\newblock {Kinetics of parametric instabilities of Alfv{\'e}n waves: Evolution
  of ion distribution functions}.
\newblock \emph{Journal of Geophysical Research (Space Physics)}, 115\penalty0
  (A9):\penalty0 A09106, September 2010.
\newblock \doi{10.1029/2009JA014987}.

\bibitem[{Gonz{\'a}lez} et~al.(2021){Gonz{\'a}lez}, {Tenerani}, {Matteini},
  {Hellinger}, and {Velli}]{Gonzalez2021}
C.~A. {Gonz{\'a}lez}, A.~{Tenerani}, L.~{Matteini}, P.~{Hellinger}, and
  M.~{Velli}.
\newblock {Proton Energization by Phase Steepening of Parallel-propagating
  Alfv{\'e}nic Fluctuations}.
\newblock \emph{ApJL}, 914\penalty0 (2):\penalty0 L36, June 2021.
\newblock \doi{10.3847/2041-8213/ac097b}.

\end{thebibliography}
\end{document}